\documentclass[letterpaper,11pt]{article}

\usepackage{dynkin-diagrams}
\usepackage{cite}
\usepackage{amsmath}
\usepackage{amssymb}
\usepackage{amsthm}
\usepackage{mathrsfs}
\usepackage{xypic}
\usepackage{tikz-cd}
\usepackage{chngcntr}
\usepackage{setspace}

\tikzcdset{scale cd/.style={every label/.append style={scale=#1},
    cells={nodes={scale=#1}}}}

\pgfkeys{/Dynkin diagram, edge length=1cm,
fold radius=.6cm, 
root-radius=.11cm,
indefinite edge/.style={
draw=black, fill=white, dotted,
thin}}

\usepackage{hyperref}
\usepackage{color}
\definecolor{darkred}{rgb}{0.65,0.15,0}
\hypersetup{pdfborder={0 0 0},colorlinks=true,urlcolor=darkred,citecolor=blue,linkcolor=darkred,linktocpage=true}

\usepackage{float} 

\usepackage{geometry}

\newgeometry{vmargin={35mm}, hmargin={35mm}} 

\usepackage{keyval}
\makeatletter
\define@key{setpar}{left}[0pt]{\leftmargin=#1}
\define@key{setpar}{right}[0pt]{\rightmargin=#1}
\define@key{setpar}{both}{\leftmargin=#1\relax\rightmargin=#1}
\makeatother

\makeatletter
\renewcommand*\env@matrix[1][\arraystretch]{%
  \edef\arraystretch{#1}%
  \hskip -\arraycolsep
  \let\@ifnextchar\new@ifnextchar
  \array{*\c@MaxMatrixCols c}}
\makeatother  

\def\eg{{\it e.g.}}
\def\ie{{\it i.e.}}

\def\DWeight#1#2#3{\bigl(\raise2.5pt\hbox{${}_{#1}$}{}^{#2}_{#3}\bigr)}

\def\AAWeight#1#2{\bigl(\raise0pt\hbox{${}^{#1}_{#2}$}\bigr)}

\def\fh{{\mathfrak h}}

\def\so{{\mathfrak{so}}}

\def\SL{{\mathrm{SL}}}
\def\Spin{{\mathrm{Spin}}}
\def\Sp{{\mathrm{Sp}}}

\def\Stab{{\mathrm{Stab}}}

\def\nn{\nonumber}
\def\so{\mathfrak{so}}

\def\*{\partial}

\def\RR{{\mathbb R}}
\def\CC{{\mathbb C}}
\def\HH{{\mathbb H}}
\def\OO{{\mathbb O}}
\def\KK{{\mathbb K}}

\numberwithin{equation}{section}
\newlength\symlength
\newlength\pluslength
\def\Re{\mathrm{Re}}
\def\Im{\mathrm{Im}}

\makeatletter
\newcommand{\oset}[3][0ex]{%
  \mathrel{\mathop{#3}\limits^{
    \vbox to#1{\kern-3\ex@
    \hbox{$\scriptscriptstyle#2$}\vss}}}}
\makeatother

\def\xprod#1{\upperset{#1}{\circ}}

\def\xprod#1{\circ\hspace{-7.5pt}\raisebox{-2.5pt}{\text{${}_{\scriptscriptstyle#1}$}}\hspace{3pt}}

\allowdisplaybreaks

\begin{document}

%\lineskiplimit=-\maxdimen
\setstretch{.99}

%{\flushright {\tt \today}\\
%{\tt \jobname.tex}\\
%}
%\vspace{1cm}

\frenchspacing

\null\vspace{-28mm}

\includegraphics[height=2cm]{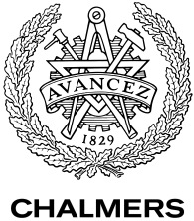}
\hspace{2mm}

\vspace{-12mm}
{\flushright Gothenburg preprint \\ 
  %\today\\
}
%March, 2021\\} %July, 2019\\}

\vspace{4mm}

\hrule

\vspace{16mm}

%{\flushright {\tt \today}\\
%{\tt linfty4.tex}\\
%}
%\null\vspace{1cm}

\thispagestyle{empty}

\begin{center}
  {\Large \bf \sc The octonionic phase space Hopf map}
    \\[10mm]
    
{\large
Martin Cederwall}

\vspace{10mm}
       {{\it Department of Physics,
         Chalmers Univ. of Technology,\\
 SE-412 96 Gothenburg, Sweden}}

%\vspace{2mm}
%       {\footnotesize ${}^2${\it Department of Mathematical Sciences,
 %       Chalmers Univ. of Technology,\\
% SE-412 96 Gothenburg, Sweden}}

\end{center}

\vfill

\begin{quote}
  
\textbf{Abstract:} 
The octonionic Hopf map, expressing $S^{15}$ as an $S^7$ bundle over $S^8$, appears in the twistor transform in 10 dimensions, $S^8$ playing the r\^ole of the celestial sphere. A symplectic lift to twistor space manifests $\Spin(2,10)$ symmetry. 
The 25-dimensional spinor orbit of $\Spin(2,10)$ is an $S^7$ bundle over the
phase space of a massless particle.
\end{quote} 

\vfill

\hrule

\noindent{\tiny email:
  martin.cederwall@chalmers.se}

\newpage

\tableofcontents

\section{Background}

It is known that a twistor transform in 10-dimensional Minkowski space \cite{Berkovits:1990yc,Cederwall:1992bi} realises the octonionic Hopf fibration,
\begin{equation}
\begin{tikzcd}[row sep = 12 pt, column sep = 12 pt]
S^7\arrow[hookrightarrow]{r}&S^{15}\ar[d]\\
&S^8
\end{tikzcd}
\label{HopfFibrationEq}
\end{equation}
the base manifold $S^8$ being the celestial sphere.
The twistor transform and its gauge symmetries use $S^7$ as {\it transformations} \cite{Cederwall:1993nx}; although not a group manifold, the parallellisable property of $S^7$ allows for vector fields generating $S^7$. The structure ``constants'' are not constants, but (some choice for) parallellising torsion.  In this sense, one can think of \eqref{HopfFibrationEq} as the appropriate generalisation of a principal bundle.
Also, it was clear that the twistor transform can be formulated in terms of a chiral spinor of $\Spin(2,10)$, the twistorial phase space \cite{Cederwall:1992bi}.

The purpose of this note is to make this statement more precise. In particular, the symplectic lift of eq. \eqref{HopfFibrationEq} (times a spectator scale) becomes the fibration
\begin{equation}
\begin{tikzcd}[row sep = 12 pt, column sep = 12 pt]
S^7\arrow[hookrightarrow]{r}&O^{25}\ar[d]\\
&\Pi^{18}
\end{tikzcd}
\label{ConfHopfFibrationEq}
\end{equation}
where the base manifold $\Pi^{18}$ is the phase space of a massless particle in 10-dimensional Minkowski space, and the total space $O^{25}$ is the 25-dimensional spinor orbit under $\Spin(2,10)$. 
In a certain sense, the existence of the 25-dimensional orbit can be attributed to the properties of $S^7$ and the octonions. 

We begin by surveying, in Section \ref{OrbitsSection}, the spinor orbits in 12 dimensions \cite{Igusa:1970}, including the constraints implementing them, their partition functions and resolutions, with focus on the 25-dimensional orbit. In Section \ref{S7TwistorSec}, we revisit the 10-dimensional twistor transform, the $S^7$ algebra, and the lift to spinors of the conformal $\Spin(2,10)$ group. Finally, the results concerning the bundle \eqref{ConfHopfFibrationEq} are collected in Section
\ref{PhaseSpaceBundleSec}.

\section{Spinor orbits under $\Spin(12)$\label{OrbitsSection}}

$\Spin(12)$ has chiral spinors of dimension $32$, which are self-conjugate symplectic.
We will mainly be interested in the real form $\Spin(2,10)$, in which spinors are real. 
Occasionally, modules of $\Spin(12)$ are denoted by Dynkin labels, \eg\ 
a chiral spinor $S=\DWeight{0000}10$. 
The classification of orbits is of course over $\CC$; we will comment later on the reality properties.

The spinor orbits in 12 dimensions were classified by Igusa \cite{Igusa:1970}.
The non-trivial orbits are of dimension 16, 25, 31 and 31, and their partial ordering forms a linear Hasse diagram.
There is a single quartic invariant\footnote{The extended algebra, in the sense of ref. \cite{CHKLS}, is $E_7$.} $I_4$ of a spinor.
Therefore, a generic orbit has dimension 31 (instead of 32), and generic orbits are labelled by the non-zero value of $I_4$. In all non-generic orbits, $I_4=0$. By a slight abuse of notation, we call the union of the generic orbits $O^{32}$, while the other orbits $O^{31}$, $O^{25}$ and $O^{16}$ are denoted by their dimension. The closure $\bar O$ of an orbit is the union of all $O_i$ such that $O_i\preceq O$ in the partial order defined by the Hasse diagram.

%Linear Hasse diagram. ($O^{32}$ means the union of the 31-dimensional orbits with non-vanishing invariant (the generic orbits)).
%Comment on ring of invariants (the single inv.  goes away in all non-generic cases).
%Constraints (are these previously known?) and partition functions for all orbits. Representatives. Really for closures $\bar O$, including all smaller (in the partial ordered sense) orbits. 

The minimal orbit $\bar O^{16}$ is the space of pure spinors $Z$, satisfying\footnote{Denote the invariant symplectic bispinor $\epsilon_{AB}$, and let $\bar Z^A=\epsilon^{AB}Z_B$, where $\epsilon^{AC}\epsilon_{CB}=-\delta^A_B$. $\Gamma_{MA'}{}^A$ are Weyl matrices. We then define the shorthand
$(XMY)=\bar X^AM_A{}^BY_B$.} $(Z\Gamma_{MN}Z)=0$, since $\DWeight{0100}00$ is the only symmetric spinor bilinear in addition to the leading one $(Z\Gamma_{M_1\ldots M_6}Z)$ in $\DWeight{0000}20$ . The next-to-generic orbit $\bar O^{31}$ is characterised by the vanishing of $I_4=(Z\Gamma_{MN}Z)(Z\Gamma^{MN}Z)$. The remaining orbit, $\bar O^{25}$, which is the one of main interest here, is characterised by the constraint $(\Gamma_{MN}Z)_A(Z\Gamma^{MN}Z)=0$.
The Hasse diagram and the constraints are collected in eq. \eqref{HasseEq}.

\begin{align}
\begin{matrix}
\bar O^{32}&&\\
\cup&&\\
\bar O^{31}&&(Z\Gamma^{MN}Z)(Z\Gamma_{MN}Z)=0\\
\cup&&\Uparrow\\
\bar O^{25}&&(\Gamma^{MN}Z)_A(Z\Gamma_{MN}Z)=0\\
\cup&&\Uparrow\\
\bar O^{16}&&(Z\Gamma_{MN}Z)=0\\
\cup&&\Uparrow\\
\{0\}&&Z_A=0
\end{matrix}
\label{HasseEq}
\end{align}

For $Z\in O^k$, denote its stabiliser group $\Stab^k$. Then,
\begin{align}
\Stab^{32}&=\SL(6)\;,\nn\\
\Stab^{31}&=\Sp(6)\ltimes\CC^{14}\;,\nn\\
\Stab^{25}&=(\Spin(7)\times SL(2))\ltimes H_{17} \;,\\
\Stab^{16}&=\SL(6)\ltimes\CC^{15}\;.\nn
\end{align}
Here, $\CC^{15}$ is an antisymmetric tensor under $SL(6)$ and $\CC^{14}$ an $\epsilon$-traceless antisymmetric tensor under $\Sp(6)$. 
The Lie algebra $\fh_{17}$ of $H_{17}$ is the Heisenberg algebra in eight dimensions:
$[t_{ai},t_{bj}]=\delta_{ab}\epsilon_{ij}t$. As a $\Spin(7)\times \SL(2)$ module, $\fh_{17}=(8,2)\oplus(1,1)$.

All orbits except the minimal one are real for $\Spin(2,10)$. The fact that $O^{16}$ is not real for this signature means that $O^{25}$ is the minimal real spinor orbit under $\Spin(2,10)$, and does not contain any singular subspace except $\{0\}$.
The real form of the stability group of $Z\in O^{25}$ then is
$(\Spin(7,\RR)\times \SL(2,\RR))\ltimes H_{17}(\RR)$.
The embedding of this group in $\Spin(2,10)$ is obtained by first considering the subgroup
$\Spin(8,\RR)\times\SL(2,\RR)\times\SL(2,\RR)\subset\Spin(2,10)$, under which
$S={\bf32}=({\bf8}_s,{\bf2},{\bf1})\oplus({\bf8}_c,{\bf1},{\bf2})$. Then, $\Spin(7,\RR)\subset\Spin(8,\RR)$ is chosen so that ${\bf8}_c={\bf1}\oplus{\bf7}$, and the second $SL(2,\RR)$ is discarded.  
Thus, ${\bf32}=2({\bf1},{\bf1})\oplus2({\bf7},{\bf1})\oplus({\bf8},{\bf2})$. Also, the vector $V={\bf12}$ decomposes as
${\bf12}=({\bf8},{\bf1})\oplus2({\bf1},{\bf2})$, and the adjoint $\wedge^2V={\bf66}$ as
${\bf66}=3({\bf1},{\bf1})\oplus({\bf1},{\bf3})\oplus2({\bf8},{\bf2})\oplus({\bf21},{\bf1})\oplus({\bf7},{\bf1})$.

%Minimal orbit $O^{16}$. Dimension 16. Stability group $\SL(6)\ltimes\CC^{15}$. Partition, constraints.
%$O^{16}$ is not real for $\Spin(2,10)$.

%Next-to-minimal orbit $O^{25}$. Dimension 25.

To each orbit $O^k$ we can associate a ring $R^k$, the coordinate ring of $\bar O^k$ generated by a spinor $Z$, where the ideal generated by the corresponding constraint $I^k=0$ in eq. \eqref{HasseEq} is divided out, \ie, $R^k=\CC[S]/\langle I^k\rangle$. This ring can be described by its Hilbert series $P^k(t)$, counting the dimensions at each degree (power of $Z$), or by its refined partition function (equivariant Hilbert series) $Z^k(t)$ taking values in the representation ring of $\Spin(12)$. We refer to the partition function of $R^k$ as the partition function of $\bar O^k$.
All the rings $R^k$ are Gorenstein.
The Hilbert series are
\begin{align}
P^{32}(t)&={1\over(1-t)^{32}}\;,\nn\\
P^{31}(t)&={1+t\over(1-t)^{31}}\;,\nn\\
P^{25}(t)&={1 + 7 t + 28 t^2 + 52 t^3 + 52 t^4 + 28 t^5 + 7 t^6 + t^7\over(1-t)^{25}}\;.\\
P^{16}(t)&={1 + 16 t + 70 t^2 + 112 t^3 + 70 t^4 + 16 t^5 + t^6\over(1-t)^{16}}\;.\nn			
\label{PoftEq}
\end{align}
The refined partition function of $\bar O^{25}$ is given by\footnote{We use the shorthand $(1-t)^{-r}=\bigoplus_{n=0}^\infty(\vee^n r)t^n$, for the partition function of the ring $\CC[r]$ freely generated by even variables in the module $r$ at degree 1.}
\begin{align}
Z^{25}(t)={1\ominus St^3\oplus\wedge^2Vt^4\ominus\tilde\vee^2Vt^6 \oplus\tilde\vee^2Vt^8\ominus\wedge^2Vt^{10}\oplus St^{11}\ominus t^{14} \over(1-t)^S} \;,
\end{align}
reflecting the minimal additive resolution over $\CC[S]$ ($\tilde\vee^2V$ means traceless symmetric product, the 77-dimensional module
$\DWeight{2000}00$).
%Let $R=R^{32}$ be the ring  freely generated by $S$ (\ie, the coordinate ring or the generic orbit). 
The minimal additive resolution of $R^{25}$,
\begin{equation}
\begin{tikzcd}[row sep = 2 pt, column sep = 10 pt, scale cd=.734]
	\CC[S]&S\otimes \CC[S]\ar[l,"d_1"]&\wedge^2V\otimes \CC[S]\ar[l,"d_2"]&\tilde\vee^2V\otimes \CC[S]\ar[l,"d_3"]
	&\tilde\vee^2V\otimes \CC[S]\ar[l,"d_4"]&\wedge^2V\otimes \CC[S]\ar[l,"d_5"]&S\otimes \CC[S]\ar[l,"d_6"]&\CC[S]\ar[l,"d_7"]
\end{tikzcd}\label{AddResEq1}
\end{equation}
has depth 7, the codimension. The explicit form of the differential is
\begin{align}
d_1\psi&=(Z\Gamma^{MN}Z)(Z\Gamma_{MN}\psi)\;,\nn\\
(d_2B)_A&=(\Gamma^{MN}Z)_AB_{MN}\;,\nn\\
(d_3T)_{MN}&=(Z\Gamma_{[M}{}^PZ)T_{N]P}\;,\nn\\
(d_4T')_{MN}&=(Z\Gamma_{(M}{}^PZ)T'_{N)P}%-{1\over12}\eta_{MN}(Z\Gamma^{PQ}Z)T'_{PQ}
	\;,\label{AddResEq2}\\
(d_5B')_{MN}&=(Z\Gamma_{(M}{}^PZ)B'_{N)P}-{1\over12}\eta_{MN}(Z\Gamma^{PQ}Z)B'_{PQ}\;,\nn\\
(d_6\psi')_{MN}&=(Z\Gamma_{MN}\psi)\;,\nn\\
(d_7\phi)_A&=(\Gamma^{MN}Z)_A(Z\Gamma_{MN}Z)\phi\;.\nn
\end{align}
In all cases, nilpotency $d_id_{i+1}=0$ follows without calculation from the absence of representations in tensor products. For example,
$d_1d_2=0$ and $d_6d_7=0$ follow from $\vee^4\DWeight{0000}10\not\supset\DWeight{0100}00$.

The $\Spin(12)$ module appearing at degree $n$ in $R^{25}$ can be given explicitly\footnote{There is a remarkable similarity with the $\Spin(11)$ spinor orbit obtained by $(\lambda\gamma^m\lambda)=0$. This ring encodes the 11-dimensional supergravity multiplet
\cite{Cederwall:2009ez,Cederwall:2010tn,Cederwall:2023wxc}. Its partition function is
$Z(t)=\bigoplus_{n=0}^\infty\bigoplus_{i=0}^{\lfloor{n\over2}\rfloor}(0,i,0,0,n-2i)t^n$. However, the dimension is 23, it is not the reduction of 
eq. \eqref{Z25asSum} to $\Spin(11)$, but a subspace of $\bar O^{25}$.} as
\begin{align}
Z^{25}(t)=\bigoplus_{n=0}^\infty\bigoplus_{i=0}^{\lfloor{n\over2}\rfloor}\DWeight{0\,i\,0\,0}{n-2i}{\;\;\;0}\,t^n\;.
\label{Z25asSum}
\end{align}

%Eq. \eqref{PoftEq} in terms of modules of stability group?

The Tate resolution of the ring $R^{25}$ is infinite\footnote{Again, there is a striking similarity to the Tate resolution of an 11-dimensional spinor with 
constraint $(\lambda\gamma^a\lambda)=0$ \cite{Cederwall:2023wxc}. There are two ``holes'' at degrees 2 and 5, and an apparent reflection symmetry, in this case around degree 2 (one may include $\so(12)$ at degree 0, and possibly prolong to negative degrees). We do not know much of the structure of the dual $L_\infty$ algebra, \eg\ whether it contains no higher brackets than a 4-bracket, or whether it is freely generated from degree 3.}. In terms of the partition function it takes the form $Z^{25}(t)=\Pi_{n=1}^\infty(1-t^n)^{-r_n}$, where the first few modules $r_n$ are
\begin{align}
r_1&=\DWeight{0000}10\;,\nn\\
r_2&=0\;,\nn\\
r_3&=-\DWeight{0000}10\;,\nn\\
r_4&=\DWeight{0100}00\;,\nn\\
r_5&=0\;,\\	
r_6&=-\DWeight{0000}00-\DWeight{0001}00-\DWeight{2000}00\;,\nn\\
r_7&=\DWeight{0000}10+\DWeight{1000}01+\DWeight{0100}10\;,\nn\\
r_8&=-\DWeight{0000}00-\DWeight{0001}00-\DWeight{0200}00\;.\nn
\end{align}

\section{$S^7$ and twistors\label{S7TwistorSec}}

Let $\KK_\nu$ be the division algebra of dimension $\nu$: $\RR$, $\CC$, $\HH$ and $\OO$ for $\nu=1,2,4,8$.
The sphere $S^\nu=\KK_\nu P^1$ can be constructed from homogeneous coordinates  $(x,y)\in\KK_\nu^2$.  
The two patches of $S^\nu$ ($y\neq0$ and $x\neq0$, respectively) have coordinates $z=xy^{-1}$ and $z'=yx^{-1}$, with overlap $z'=z^{-1}$.
When $\KK_\nu$ is associative, $\KK_\nu P^1$ is obtained by an equivalence relation 
$(x,y)\approx(xa,ya)$, $a\in\KK_\nu$, since this transformation preserves $z$: $(xa)(ya)^{-1}=(xa)(a^{-1}y^{-1})=xy^{-1}$.
For the non-associative octonions, this does not work. If $x\mapsto xa$, one needs $y\mapsto (yx^{-1})(xa)$, since then
$yx^{-1}\mapsto ((yx^{-1})(xa))(xa)^{-1}=yx^{-1}$. In ref. \cite{Cederwall:1993nx}, this was referred to as the ``$x$-product'',
\begin{align}
y\mapsto y\xprod xa\equiv(yx^{-1})(xa)\;.
\end{align}
This product, for a given $x$, has all the properties of the standard octonionic product. It defines $S^7$ transformations in a consistent way, for $|a|=1$.
The Moufang identity and alternativity of the associator imply $a\xprod{x}b=x^{-1}((xa)b)=(ax^{-1})(xb)=(a(bx^{-1}))x$.
One can equally well start from $y\mapsto yb$, the relation between $a$ and $b$, 
$b=x^{-1}((xy^{-1}(yb))=x^{-1}(x\xprod y b)=y^{-1}\xprod x (yb)$ 
is bijective.

Consider repeated transformations. The variable $x$ defining the product itself transforms. Let $R_ax=xa$, $R_ay=y\xprod x a$.
Then, 
\begin{align}
R_aR_bx&=(xa)b=x(a\xprod x b)\;,\nn\\
R_aR_by&=(y\xprod x a)\xprod{\hspace{-1.5pt}xa}b=y\xprod x(a\xprod xb)\;.
\end{align}
Group-like composition of $S^7$ transformations is obtained as $a\xprod xb$.

Let 
\begin{align}
\lambda_\alpha=\left(\begin{matrix}x\\y\end{matrix}\right)
\end{align}
be a spinor of $\Spin(9,\RR)$ in 2-component octonionic notation \cite{Sudbery_1984,Cederwall:1992bi,Cederwall:1993xe}.
There is an invariant norm, and we can restrict to $S^{15}$: $\lambda^\dagger\lambda=|x|^2+|y|^2=1$.
$\Spin(9,\RR)$ vectors are represented (in an octonionic gamma matrix basis) as $2\times2$ hermitean traceless matrices. 
Real and imaginary parts are defined as $\Re x={1\over2}(x+\bar x)$, $\Im x={1\over2}(x-\bar x)$.
The spinor bilinear
\begin{align}
v=\lambda\lambda^\dagger-{1\over2}(\lambda^\dagger\lambda)I
=\left(\begin{matrix}{1\over2}(|x^2|-|y|^2)&x\bar y\\ y\bar x&{1\over2}(|y^2|-|x|^2)\end{matrix}\right)
\end{align}
is a unit vector: $v^2\equiv4\det v=(|x^2|-|y|^2)^2+4|x|^2|y|^2=(|x|^2+|y|^2)^2=1$.
Let $\omega$ be a spinor with components $p,q$ conjugate to $\lambda$ with Poisson bracket $\{\omega^\alpha,\lambda_\alpha\}=\delta^\alpha_\beta$.
Generators (through Poisson bracket) of $S^7$ can be written either
$W_x=\Im(\bar xp+\bar y\xprod xq)$ or $W_y=\Im(\bar x\xprod yp+\bar yq)$. Namely, consider for example
\begin{align}
W_x(a)\equiv\Re(\bar aW_x)=\Re((\bar px)a+(\bar q\xprod xy)a)=\Re(\bar p(xa)+\bar q(y\xprod xa))\;;.
\end{align}
for a parameter $a\in\OO$ with $\Re\,a=0$.
 Then, $\{W_x(a),x\}=xa$, $\{W_x(a),y\}=y\xprod xa$.
This provides a concrete construction of the Hopf map, and indeed of $S^{15}$ as a ``principal $S^7$ bundle'' over $S^8$; it is meaningful to write
$S^8=S^{15}/S^7$.

Consider the spinor generators
\begin{align}
T=(\lambda\lambda^\dagger)\omega-\lambda\Re(\lambda^\dagger\omega)\;.\label{TGenEq}
\end{align}
Explicit evaluation gives
\begin{align}
T=\left(\begin{matrix}x\Im(\bar xp+\bar y\xprod xq)\\y\Im(\bar x\xprod yp+\bar yq)\end{matrix}\right)\;.
\end{align}
showing that $T$ contains only 7 independent generators.

The spinor form \eqref{TGenEq} of the generators is covariant under $\Spin(1,9)$ (provided of course that the constraint $\lambda^\dagger\lambda=1$ is relaxed).
The twistor transform in $d=10$ Minkowski space reads
\begin{align}
p&=\lambda\lambda^\dagger\;,\nn\\
\omega&=x\lambda\;,
\end{align}
or, in standard real notation
\begin{align}
p_a&={1\over2}(\lambda\gamma_a\lambda)\;,\nn\\
\omega^\alpha&=x^a(\gamma_a\lambda)^\alpha\;.\label{10TwistorEq}
\end{align}
From eq. \eqref{10TwistorEq}, it follows, using the Fierz identity $(\gamma^a\lambda)^\alpha(\lambda\gamma_a\lambda)=0$, that
\begin{align}
{1\over2}(\lambda\gamma^a\lambda)(\gamma_a\omega)_\alpha-\lambda_\alpha(\lambda\omega)
={1\over2}(\lambda\gamma^a\lambda)(\hspace{-10pt}\underbrace{\gamma_a\gamma_b}_{2\eta_{ab}-\gamma_b\gamma_a}\hspace{-10pt}\lambda)_\alpha x^b
	-\lambda_\alpha(\lambda\gamma_a\lambda)x^a
=0\;.
\end{align}
Indeed,
\begin{align}
T_\alpha={1\over2}(\lambda\gamma^a\lambda)(\gamma_a\omega)_\alpha-\lambda_\alpha(\lambda\omega)
\end{align}
is identical to eq. \eqref{TGenEq}, in standard notation.
The Hopf fibration becomes equipped with a spectator scale $\RR_+$. $S^8$ is the celestial sphere, and $S^8\times\RR_+$ is the forward light-c\^one.

The reducibility \cite{Berkovits:1990yc} of the generators $T_\alpha$ follows as
$(\lambda\gamma_a T)=0$. Then, any vector $v_a=(\lambda\gamma_a\mu)$ satisfies $(\lambda\gamma^a\lambda)v_a=0$, so there is a scalar syzygy. 
The resolution of this system, in the form of a differential (BRST operator), was given in ref. \cite{Cederwall:1992bi} and rediscovered in component formalism in ref. \cite{Nahari:2024}.
In total, the number of  independent components in $T_\alpha$ is $16-10+1=7$.
The Poisson brackets between generators are
\begin{align}
\{T_\alpha,T_\beta\}=\lambda_\alpha T_\beta-\lambda_\beta T_\alpha\;.
\end{align}

A massless particle in 10-dimensional Minkowski space is conformally symmetric.
The spinors $\lambda_\alpha$ and $\omega^\alpha$ are combined in the $\Spin(2,10)$ chiral spinor
\begin{align}
Z_A=\left(\begin{matrix}\lambda_\alpha\\\omega^\alpha\end{matrix}\right)\;.
\end{align}
Direct calculation, using the decomposition of $\Spin(2,10)$ $\Gamma$-matrices in $\Spin(1,9)$ $\gamma$-matrices gives
\begin{align}
J_A\equiv{1\over24}(\Gamma_{MN}Z)_A(Z\Gamma^{MN}Z)
=\left(\begin{matrix}T_\alpha\\U^\alpha\end{matrix}\right)\;,
\end{align}
where $U^\alpha=-{1\over2}(\omega\gamma^a\omega)(\gamma_a\lambda)^\alpha+\omega^\alpha(\omega\lambda)$.
It follows from the twistor transform that also $U^\alpha=0$, so $J_A$ should only have 7 independent components. 
This is shown in Section \ref{OrbitsSection}.
The Poisson bracket between $T$ and $U$ becomes, after a somewhat difficult calculation,
\begin{align}
\{T_\alpha,U^\beta\}=\lambda_\alpha U^\beta-\omega^\beta T_\alpha+((\lambda U)-(\omega T))\delta_\alpha^\beta\;.
\end{align}
All brackets are collected in the $\Spin(2,10)$-covariant statement
\begin{align}
\{J_A,J_B\}=Z_AJ_B-Z_BJ_A-{1\over2}\epsilon_{AB}(ZJ)\;.\label{JJEq}
\end{align}

\section{The phase space Hopf map\label{PhaseSpaceBundleSec}}

The considerations in Section \ref{S7TwistorSec} and ref. \cite{Cederwall:1992bi} show that the phase space of a massless particle in 10-dimensional Minkowski space is described by a conformal spinor $Z_A$, self-conjugate with respect to the Poisson bracket, $\{Z_A,Z_B\}=\epsilon_{AB}$, with the first class constraints
$J_A={1\over24}(\Gamma_{MN}Z)_A(Z\Gamma^{MN}Z)=0$.
The constraint surface is the 25-dimensional $\Spin(2,10)$ spinor orbit $O^{25}$ (the origin excluded), with
an additive resolution given in eqs. \eqref{AddResEq1} and \eqref{AddResEq2}.
The physical phase space $\Pi^{18}$  is the space of $S^7$ orbits (generated by $J$) in $O^{25}$.
Note that $O^{25}$, as well as all other spinor orbits, is preserved by $S^7$. The Lorentz generators, proportional to $(Z\Gamma_{MN}Z)$ are expressible in terms of $x$ and $p$, so invariant under $S^7$ (this is however not an issue, since $O^{16}$ is not real). The constraint surface $J_A=0$ is preserved thanks to eq. \eqref{JJEq}.
There are no fixed points except 0.
The cotangent/symplectic lift of the Hopf map (with a spectator scale)
\begin{equation}
\begin{tikzcd}[row sep = 12 pt, column sep = 12 pt]
S^7\arrow[hookrightarrow]{r}&S^{15}\times\RR_+\ar[d]\\
&S^8\times\RR_+
\end{tikzcd}
%\label{HopfFibrationEq}
\end{equation}
becomes 
\begin{equation}
\begin{tikzcd}[row sep = 12 pt, column sep = 12 pt]
S^7\arrow[hookrightarrow]{r}&O^{25}\ar[d]\\
&\Pi^{18}
\end{tikzcd}
%\label{ConfHopfFibrationEq}
\end{equation}
which is a phase space version of the Hopf map. In both cases, the base manifold is constructed as the quotient by $S^7$ transformations,
$\Pi^{18}=O^{25}/S^7$.

\newpage

\bibliographystyle{utphysmod2}
%\bibliography{biblio}
%\nocite{*}

\providecommand{\href}[2]{#2}\begingroup\raggedright\endgroup

\end{document}